\documentclass[reprint,amsmath,amssymb,aps,prl,superscriptaddress]{revtex4-2}

\usepackage{graphicx}
\usepackage{dcolumn}
\usepackage{bm}
\usepackage[margin=2.0cm]{geometry}
\usepackage{amsmath}
\usepackage{braket}
\usepackage{appendix}
\usepackage{float}
\usepackage{siunitx}
\usepackage{amssymb}
\usepackage{color}
\usepackage{multirow}
\usepackage{longtable}
\usepackage{ulem}
\usepackage{soul}

\begin{document}

\title{ Analytically controlling laser-induced electron phase in sub-cycle motion
}
\author{Doan-An Trieu}
    \affiliation{Computational Physics Key Laboratory K002, Department of Physics, Ho Chi Minh City University of Education, Ho Chi Minh City 72711, Vietnam}
\author{Trong-Thanh D. Nguyen}
    \affiliation{Computational Physics Key Laboratory K002, Department of Physics, Ho Chi Minh City University of Education, Ho Chi Minh City 72711, Vietnam}
\author{Thanh-Duy D. Nguyen}
    \affiliation{Computational Physics Key Laboratory K002, Department of Physics, Ho Chi Minh City University of Education, Ho Chi Minh City 72711, Vietnam}
 \author{Thanh Tran}
    \affiliation{Computational Physics Key Laboratory K002, Department of Physics, Ho Chi Minh City University of Education, Ho Chi Minh City 72711, Vietnam}
\author{Van-Hoang Le}
    \affiliation{Computational Physics Key Laboratory K002, Department of Physics, Ho Chi Minh City University of Education, Ho Chi Minh City 72711, Vietnam} 
\author{Ngoc-Loan Phan}
    \email{loanptn@hcmue.edu.vn}
    \affiliation{Computational Physics Key Laboratory K002, Department of Physics, Ho Chi Minh City University of Education, Ho Chi Minh City 72711, Vietnam}

\date{\today}

\begin{abstract}
 
Precise control of the electron phase accumulated during its sub-cycle motion within intense laser fields is essential in strong-field physics, yet remains mostly indirect and complicated so far. In this Letter, we develop a novel approach to control this sub-cycle electron phase by tuning a low-frequency electric field applied on a centrosymmetric gaseous target during its interaction with a few-cycle infrared laser pulse. Our method is based on a universal analytical relation between the low-frequency electric field and its induced harmonic frequency shift, derived by the strong-field approximation. This simple relation and its universality are confirmed numerically by directly solving the time-dependent Schr\"odinger equation. Moreover, we discuss the benefits of the discovered relation in \textit{in situ} applications, including continuously and precisely tuning XUV waves and developing a new method of comprehensively sampling THz pulse. 

\end{abstract}

\maketitle

\textit{Introduction ---}   
The high-order harmonic generation (HHG) process serves as a well-established tool for probing the laser-target dynamics. One interesting aspect is the laser-target symmetry-breaking manifesting through the odd-even harmonics \cite{Frumker:PRL12,Luu:PRB16,Langer:NP17,Nguyen:pra22,Shafir:NatPhys09,Kim:NatPho14,Li:PRA20,Li:NC23,Trieu:PRA23,Vampa:NP18,Dudovich:NatPhys:06,Shafir:Nat12,He:NatCom22} or the shift of harmonic frequency~\cite{Zhou:prl96,Graml:PRB23,Bian:prl14,Lan:PRL17,He:NC18,He:prl18,You:OL17,Schmid:Nat21,Naumov:PRA15,Mandal:OE22,Zheng:OE23,Schubert:NP14}. Essentially, the underlying physics of these intensity modulations and frequency shifts are hinted deeper, originating from the interference of attosecond bursts emitted at each electron recombination event after completing a closed classical sub-cycle trajectory in a laser field following the ionization~\cite{Corkum:prl93, Lewenstein:pra94}. 
Central to this phenomenon is the phase disparity among adjacent attosecond bursts, which for symmetric targets, is closely linked to the electron-wave's phases accumulated during the electron quasi-classical motion within laser fields~\cite{Corkum:prl93,Lewenstein:pra94}. These cues suggest strategies of leveraging measured HHG to \textit{in situ} control electron phase (the core of strong-field laser-matter interaction) through precisely controlling sub-cycle electron trajectories within sub-femtosecond resolution via tuning the interacting laser fields~\cite{Li:PRA20,Li:NC23,Trieu:PRA23,Vampa:NP18,Dudovich:NatPhys:06,Shafir:Nat12,Kim:NatPho14,He:NatCom22,Zhou:prl96,Graml:PRB23,Bian:prl14,Lan:PRL17,Naumov:PRA15,He:NC18,You:OL17,Schmid:Nat21,Zheng:OE23,Mandal:OE22,He:prl18,Schubert:NP14}.

In the past decades, various ways of laser tuning have been intensively proposed, including compressing laser pulse duration to a few optical cycles~\cite{Bian:prl14,Lan:PRL17,Naumov:PRA15,He:NC18,You:OL17,Schmid:Nat21,He:prl18}, adjusting the chirp of the driving pulse~\cite{Zhou:prl96,Graml:PRB23}, or implementing additional fields~\cite{Li:NC23,Li:PRA20,Kim:NatPho14,Shafir:Nat12,Trieu:PRA23,He:NatCom22,Vampa:NP18,Dudovich:NatPhys:06,Bao:PRA96,Wang:JPB98,Hong:JPB07}. Controlling electron  phases allows resolving the quantum ionization and recombination time~\cite{Shafir:Nat12}, \textit{in situ} characterizing attosecond bursts~\cite{Dudovich:NatPhys:06,You:OL17}, tuning XUV waves~\cite{Mandal:OE22,Naumov:PRA15,Zhou:prl96}, sampling waveform of broadband fields~\cite{Li:PRA20,Vampa:NP18,Trieu:PRA23,Li:NC23}, measuring structures~\cite{Schmid:Nat21,Zheng:OE23}, or probing ultrafast dynamics of targets~\cite{He:NC18,He:NatCom22,He:prl18,Lan:PRL17,Bian:prl14}.
However, these HHG-based controls of sub-cycle electron  phases are still complex or indirect. The reason is the absence of an explicit analytical form connecting the phases of electron waves with the HHG measurements. Constructing such a relation is not easy due to the nonlinearity of the considered effects.

This Letter establishes a novel analytical framework directly relating harmonic frequency shifts with the intensity of a low-frequency electric field introduced alongside a few-cycle driving laser pulse during interaction with atomic or centrosymmetric molecular gases. Consequently, it allows for the direct and precise manipulation of the phases of sub-cycle electron waves by monitoring the harmonic frequency shift while tuning the low-frequency electric field. The direct and simple relation is also confirmed numerically by solving the time-dependent Sch{\"o}dinger equation. Remarkably, the analytical relationship is universal regardless of targets or driving laser parameters. We then discuss its application in \textit{in situ} tuning XUV waves and THz metrology.

\begin{figure}[hbt!]
    \begin{center}
    \includegraphics[width=1.0\linewidth]{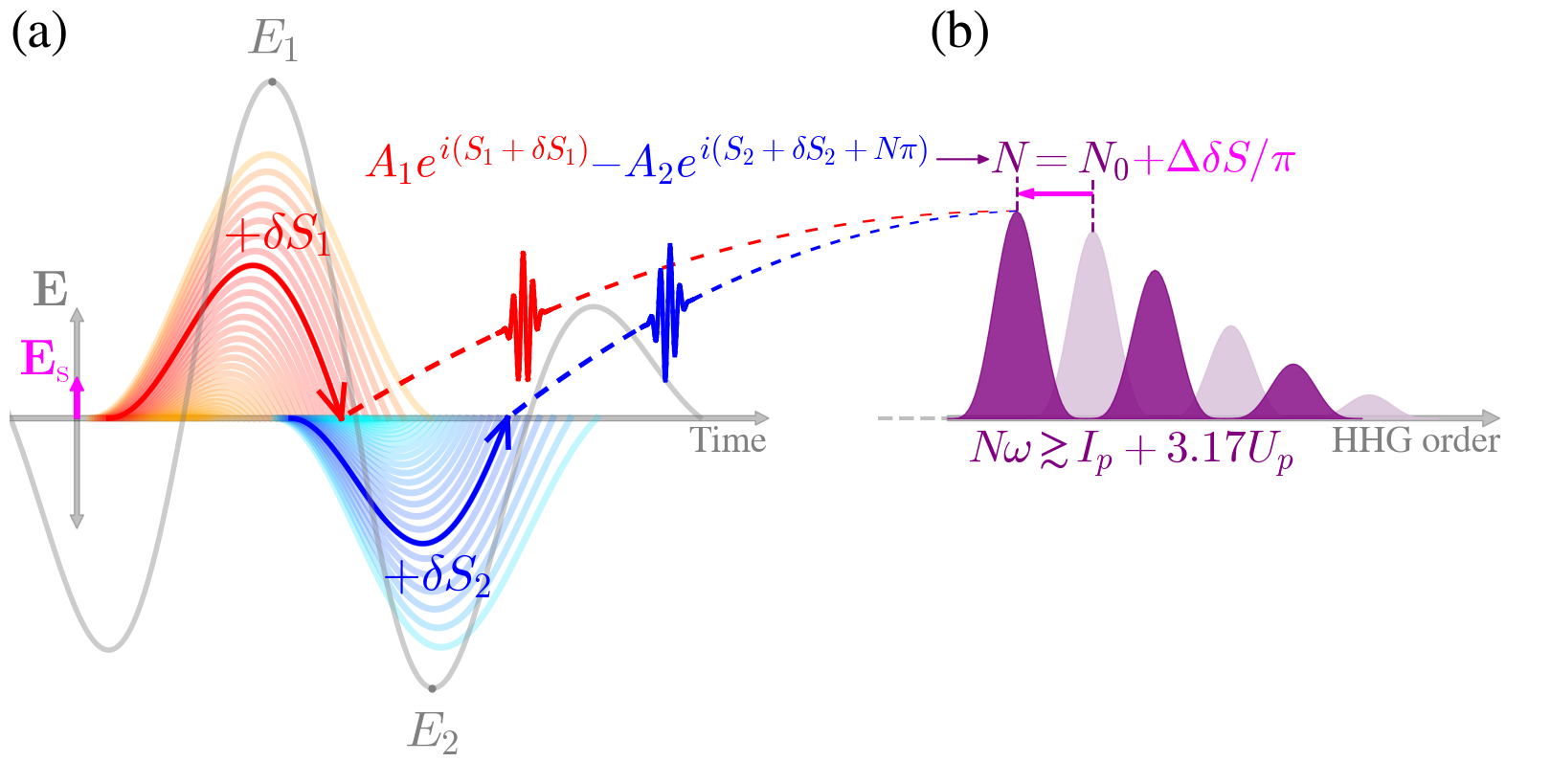}
    \end{center}
    \caption{Transformation from (a)~SEF-induced changes of electron phases ($\delta S_1$ and $\delta S_2$) in the time domain into (b)~SEF-induced frequency shifts of harmonics (from order $N_0$ to $N$) in the frequency domain. The added static field $E_s$ perturbs the actions of the two adjacent electron trajectories ($S_1$ and $S_2$), leading to the distortion of emission bursts ($A_1 e^{i S_1}$ and $A_2 e^{i S_2}$), resulting in shifting their interference pattern by ${\Delta \delta S}/{\pi}$, where the phase distortion difference $\Delta\delta S=\delta S_1-\delta S_2$. }
    \label{fig:mechanism}
\end{figure}

{\textit{Analytical relation} --- } Before deriving an analytical formula describing harmonic frequency shift induced by adding a low-frequency electric field to the driving laser pulse, we first review the harmonic generation from a centrosymmetric target interacting with a few-cycle laser pulse in the recollision picture~\cite{Corkum:prl93,Lewenstein:pra94}. 

In the time domain, HHG results essentially from the interference of emission bursts radiated every half-cycle intervals when ionized electrons recombine with the parent ion after being quasi-classically driven by the laser electric field. Employing a few-cycle driving laser pulse ensures that no more than two emission bursts called $A_1 e^{-i \phi_1}$ and $A_2 e^{-i \phi_2}$ (at the middle of the pulse) contribute to generating harmonics near the cutoff, making their interference pattern much more apparent. In the recollision picture~\cite{Corkum:prl93,Lewenstein:pra94}, the phase of the emission burst is $\phi = N_0 \omega_0 t_r - S(t_i, t_r, p)$, where $S(t_i, t_r, p)$ is the quasi-classical action of the electron trajectories launched at ionization instant $t_i$ and finished by the recombination at $t_r$; $p$ is the canonical momentum; $\omega_0$ is the carrier frequency of driving laser; and $N_0$ is the harmonic order. The interference of the two bursts $A_1 e^{-i \phi_1} - A_2 e^{-i \phi_2}$ gives the interference pattern with a maximum in the order $N_0 = (2 k +1) + \Delta S / \pi$, where $k$ is an integer number, and $\Delta S = S_1 - S_2$ is the phase difference. In the case of a few-cycle laser pulse, harmonic peaks may deviate from the odd-only pattern due to the non-vanishing $\Delta S$ caused by the nonidentical electric fields felt at the two adjacent half-cycles.

Now we discuss the effect of adding a weak low-frequency electric field, which can be adiabatically considered as a static electric field (SEF), into a few-cycle driving laser. Figure~\ref{fig:mechanism} sketches the mechanism. (i)~A weak SEF deforms electron trajectories, leading to a change in the quasiclassical actions and, as a consequence, distorting the phase of attosecond bursts. The corresponding phase distortions are labeled as $\delta S_1$ and $\delta S_2$ for the two bursts in the middle of the pulse. (ii) The SEF-induced phase distortion shifts the interference pattern by order of $ \Delta N \equiv N - N_0 = {\Delta~\delta S} / {\pi}$, where $\Delta\delta S = \delta S_1 - \delta S_2$. 

To treat this SEF-induced harmonic frequency shift analytically, we need to express the SEF-induced phase distortion difference $\Delta\delta S$ in an explicit mathematical form. To do this, we represent the instantaneous total electric field as $\mp E_i \sin \omega_0 t + E_s$, where $E_i$ is the peak amplitude of the driving laser at the half-cycle responsible for the interested emission burst, and $E_s$ is the SEF. Applying the strong-field approximation, we can express the SEF-induced phase distortion as
\begin{equation}
    \delta S_i = \pm C(N_0) \dfrac{E_i}{\omega_0^3} E_s,
    \label{eq:phase_shift}
\end{equation}
where $C(N_0) = 2 (\sin \theta)(\Delta \theta \cos \Delta \theta - \sin \Delta \theta)$ is a dimensionless coefficient, with $\theta = \omega_0 (t_r + t_i)/2$ and $\Delta \theta =  \omega_0 (t_r - t_i)/2$. The detailed proof is presented in Supplementary~\cite{suppl}. Although we derived a similar formula in Ref.~\cite{Trieu:PRA23}, it is limited to the case of multi-cycle driving lasers only. Here, the derived Eq.~(\ref{eq:phase_shift}) is more general and can be applied to any pulses.  

From Eq.~(\ref{eq:phase_shift}), we obtain frequency shifts of harmonic peaks in an analytical form
\begin{equation}
    \Delta N = \pm C(N_0) \dfrac{E_m}{\pi \omega_0^3} E_s,
    \label{eq:ana}
\end{equation}
where $E_m=(E_1+E_2)/2$ is the average peak amplitude of two adjacent half-cycles of a few-cycle driving field. The symbol $\pm$ implies the blueshift or redshift. Specifically, a blueshift occurs if the first attosecond burst emits at the half-cycle where the electric field of the driving laser and the SEF are in opposite directions.  Meanwhile, the second attosecond burst emerges when the two electric fields are in the same direction. Conversely, a redshift happens. 

It should be noted that in Eq.~(\ref{eq:ana}), we approximate $C(N)\approx C(N_0)$ because $\Delta N \ll N_0$ near the cutoff. In addition, in weak perturbation $E_s \ll E_m$, we assume the SEF almost does not alter ionization and recombination instants for each harmonic. Therefore, the coefficient $C(N)$ is universal, i.e., independent of the laser parameters, and can easily be evaluated using the saddle-point approximation~\cite{Lewenstein:pra94,Salieres:Sci01}. Within this approximation, the coefficient for harmonics near the cutoff is $C = 2.558$. 

In summary, we analytically discover that harmonic frequency shifts occur by and linearly change with varying the low-frequency electric field in the perturbation regime. These harmonic frequency shifts reflect the SEF-induced phase distortion of the electron phase in the time domain. The rule governing the SEF-induced frequency shifts is universal, meaning they are independent of the target parameters as long as the target is centrosymmetric.

\begin{figure*} [t!]
    \begin{center}
        \includegraphics[width=0.95\linewidth]{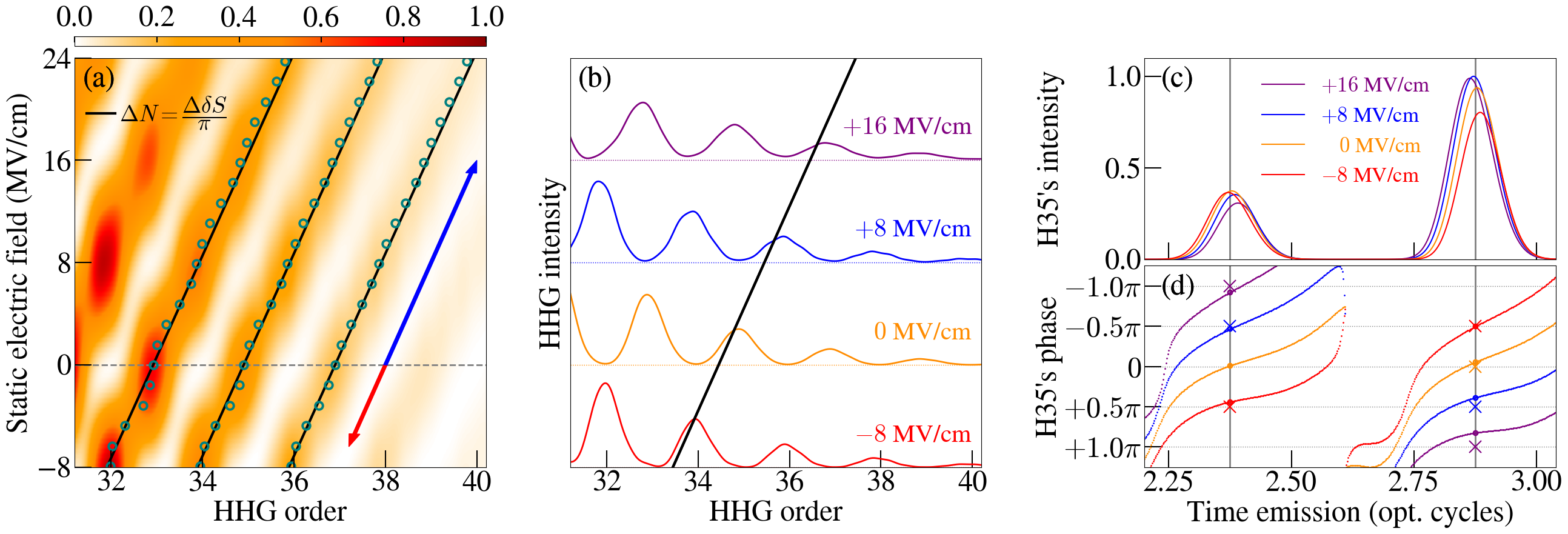}
    \end{center}
    \caption{Consistency between the numerical simulation and analytical prediction of SEF-induced harmonic frequency shift (in frequency domain)~(a)-(b), and the SEF-induced phase distortion of the attosecond bursts (in time domain)~(d) for harmonics near the cutoff. The simulation is conducted by solving the TDSE for a hydrogen atom in a combination of a SEF and a five-cycle sine-squared pulse with an intensity of $2\times 10^{14}$~W/cm$^2$, a wavelength of 800~nm, and a CEP of $-\pi/2$. For a convenient presentation, HHG intensity in Fig.~(a) is normalized to its maximum. Open circles pick the centroids of simulated harmonic peaks, which indicate clear redshift ($E_s < 0$) and blueshift ($E_s > 0$) as predicted analytically by Eq.~(\ref{eq:ana}) (black straight lines). Figure~(b) presents HHGs for specific SEF intensities when harmonic peaks shift by $-1$, $0$, $+1$, and $+2$ orders. Figure~(d) shows the phases of 35$^\mathrm{th}$-order's attosecond bursts at their emission instants, defined by the peaks of the time profile's intensity (grey vertical lines)~(c). The simulated SEF-induced phase distortions are well consistent with analytical predictions by Eq.~(\ref{eq:phase_shift}), denoted by crosses in Fig.~(d). 
   }    
    \label{fig:numerical}
\end{figure*}

{\textit{Numerical validation} --- } We validate the obtained analytical rule of SEF-induced harmonic frequency shifts [Eq.~(\ref{eq:ana})] by comparing it with numerical simulation. To this end, we numerically solve the time-dependent Schr\"odinger equation (TDSE) of a hydrogen atom exposed to external electric fields as
\begin{equation}
    i\dfrac{\partial}{\partial t}\Psi(\bold{r},t)=\left[ 
-\dfrac{\Delta}{2} + V_c(r)  + \bold{r} \bold{E}(t)  \right] \Psi(\bold{r},t),
\end{equation}
in which $V_c(r)$ is the ion-electron potential of the hydrogen atom, $\bold{E}(t)$ is the total electric field synthesized from a SEF and a few-cycle laser pulse. $\bold{E}(t)=\bold{\hat{e}}~ [E_0\sin^2(\pi t/\tau)\cos(\omega_0 (t-\tau/2) + \alpha) + E_s]$, where $E_0$ is the peak amplitude, $\alpha$ is the carrier-envelope phase (CEP), $\tau$ is the pulse duration, and $\bold{\hat{e}}$ is a unit vector chosen along $z$-axis. 

We use the OCTOPUS source code~\cite{Marques:CPC06} to solve the three-dimensional (3D) TDSE and the split operator method~\cite{Baudrauk:CPL91} to solve the one-dimensional (1D) TDSE with the soft-Coulomb potential $V_c(z) = - 1/\sqrt{1+4 z^2}$~\cite{Majorosi:PRA18}. The results for the 1D case, including Fig.~\ref{fig:numerical}, fully agree with those of the 3D simulation (see them in Supplementary~\cite{suppl}). Therefore, we use only 1D calculations for different laser parameters and application evidence (shown in the next section) to save computation resources. It is reasonable because the electron wave packet mostly spreads in the polarization direction of the linearly polarized laser fields. Besides, our 1D use is consistent with Ref.~\cite{Majorosi:PRA18}, which confirms that the used 1D density-based potential $V_c(z)$ can generate HHG that qualitatively matches those from the full 3D simulation.  

High-order harmonics near the cutoff, calculated by numerically solving the TDSE with varied SEF intensity and a five-cycle driving laser pulse, are shown in Figs.~\ref{fig:numerical}~(a)-(b) compared with the analytical prediction by formula~\eqref{eq:ana}. The laser parameters are 
an intensity of $2\times 10^{14}$~W/cm$^2$, a wavelength of 800~nm, and a CEP of $-\pi/2$. For clearer navigation of harmonics as varying SEF, we pick the centroids of harmonic peaks and mark them with open teal circles. Figure~(b) depicts Fig.~(a) slices at specific SEF intensities in four cases, which cause one-order redshift (i), without shift (ii), and one-order (iii) and two-order (iv) blueshifts. The figures reveal that as the SEF intensity varies, the numerically simulated harmonic peaks linearly shift and well match the analytical prediction presented by the black straight lines. Specifically, the shifts are significant, of one harmonic order for each small SEF change of about 8~MV/cm, which makes it easy to confirm experimentally. 

As discussed above, the SEF-induced frequency shift in the frequency domain is indeed the consequence of the difference in the phase distortion of adjacent attosecond bursts caused by the SEF-perturbed electron trajectories in the time domain. Therefore, we perform the validation at a deeper level, examining the analytical phase distortion~[Eq.~(\ref{eq:phase_shift})] by comparing the phases of the two emission bursts at emission instants retrieved from the Gabor transformation of the numerical HHG spectra. Results are shown in Figs.~\ref{fig:numerical} (c)-(d) and analyzed in the following.   

Figure~\ref{fig:numerical}(d) plots the 35 harmonic order (H35) phases at different instants with varied SEF intensity at different instants. The emission instants are determined by the peaks of the time profile's intensity, as presented in Fig.~\ref{fig:numerical}(c). It shows that H35 emits at two instants, about $2.37$ and $2.87$ $T_0$ ($T_0$ is an optical cycle), almost unchanged with varying the SEF. It confirms the weak-perturbation approximation used for the analytical formula~\eqref{eq:ana} that the SEF almost unaffected the recombination instants. Meanwhile, the SEF remarkably distorts the emission phase, as shown in Fig.~\ref{fig:numerical}(d). Moreover, the numerically calculated difference of SEF-induced phase distortions between the two emission bursts is fully consistent with those calculated analytically by Eq.~\eqref{eq:ana}, denoted by crosses in Fig.~\ref{fig:numerical}(d). Indeed, the SEF intensities of $-8$, $0$, $8$, and $16$ MV/cm changes the phase differences by $-1\pi$, $0$, $+1 \pi$, and $+2 \pi$. The corresponding predicted SEF-induced harmonic frequency shift by analytical formula~\eqref{eq:ana} are $-1$, $0$, $+1$, and $+2$ orders, consistent with the numerical simulation in Figs.~\ref{fig:numerical} (a)-(b).

Furthermore, we justify the analytical formula~\eqref{eq:ana} for various laser-target systems. Detailed numerical evidence is presented in Supplementary~\cite{suppl}. Particularly, we demonstrate the good consistency of the analytically predicted harmonic frequency shift with those numerically simulated (i) using various few-cycle driving lasers with a broad range of laser parameters -- intensity ranging in $[1,~3]\times 10 ^{14}$~W/cm$^2$, wavelength in [$800,~2000$]~nm, CEP within $[-34 \pi/36,~-4\pi/36]$ and $[2 \pi/36,~32\pi/36]$. These ranges of laser parameters are chosen to ensure well-resolved harmonic peaks. Besides, a laser duration of less than seven optical cycles is utilized to avoid complex interference from many emission bursts for harmonics near the cutoff. This good consistency is also obtained for (ii) different symmetric gaseous targets -- atoms and centrosymmetric molecules. Finally, the numerical simulation validates the analytical rule [Eq.~\eqref{eq:ana}] (iii) for high-energy harmonics below the cutoff under good experimental phase-matching conditions. 

We emphasize that SEF-induced harmonic frequency shifts are feasible to observe experimentally. First, the CEP stabilization for a five-cycle laser pulse (as used in this paper) is achievable with conventional technique~\cite{Haworth:NP07}. Second, currently available techniques can precisely measure the laser's peak intensities with errors of about 10\% or lower \cite{Gibson:PRL04,Alnaser:PRA04,Smeenk:PRA11,Pullen:PRA13,Wang:PRL19}, leading to minor errors of harmonic frequency shifts. Finally, detecting frequency shifts is entirely possible with regular spectrometers~\cite{Wunsche:RSI19}.

\textit{Applicability --- } Thanks to the analytical relation of SEF-induced frequency shift, we propose that it can be applied in various aspects of \textit{in situ} probing dynamics through HHG-based controlling the electron phase during sub-cycle motion by using an additional weak low-frequency field.

\noindent{\textit{(1) Continuously tuning XUV harmonics:} } HHG is an effective source of XUV pulses, essential in many applications in attosecond-resolved spectroscopy~\cite{Uiberacker:Nat7,Tzallas:NatPhys11,Sie:NatCom19} and photoelectron spectroscopy~\cite{Sie:NatCom19,Fabris:NatCom15}. For these applications, tuning parameters to obtain precise frequency and CEP of certain XUV harmonics is necessary. One of the practical methods is varying the frequency shift of the harmonics by using chirped lasers~\cite{Zhou:prl96,Graml:PRB23}, two-color laser fields~\cite{Shafir:Nat12,Dudovich:NatPhys:06}, few-cycle laser pulses~\cite{Naumov:PRA15,He:prl18,You:OL17,Schmid:Nat21}, or double infrared pulses~\cite{Mandal:OE22}. With the derived analytical formula~(\ref{eq:ana}), we propose that manipulating harmonic frequency shift can be performed analytically by tuning the low-frequency field when centrosymmetric targets interact with a few-cycle laser pulse.     

\begin{figure} [htb!]
    \centering
    \includegraphics[width=0.95\linewidth]{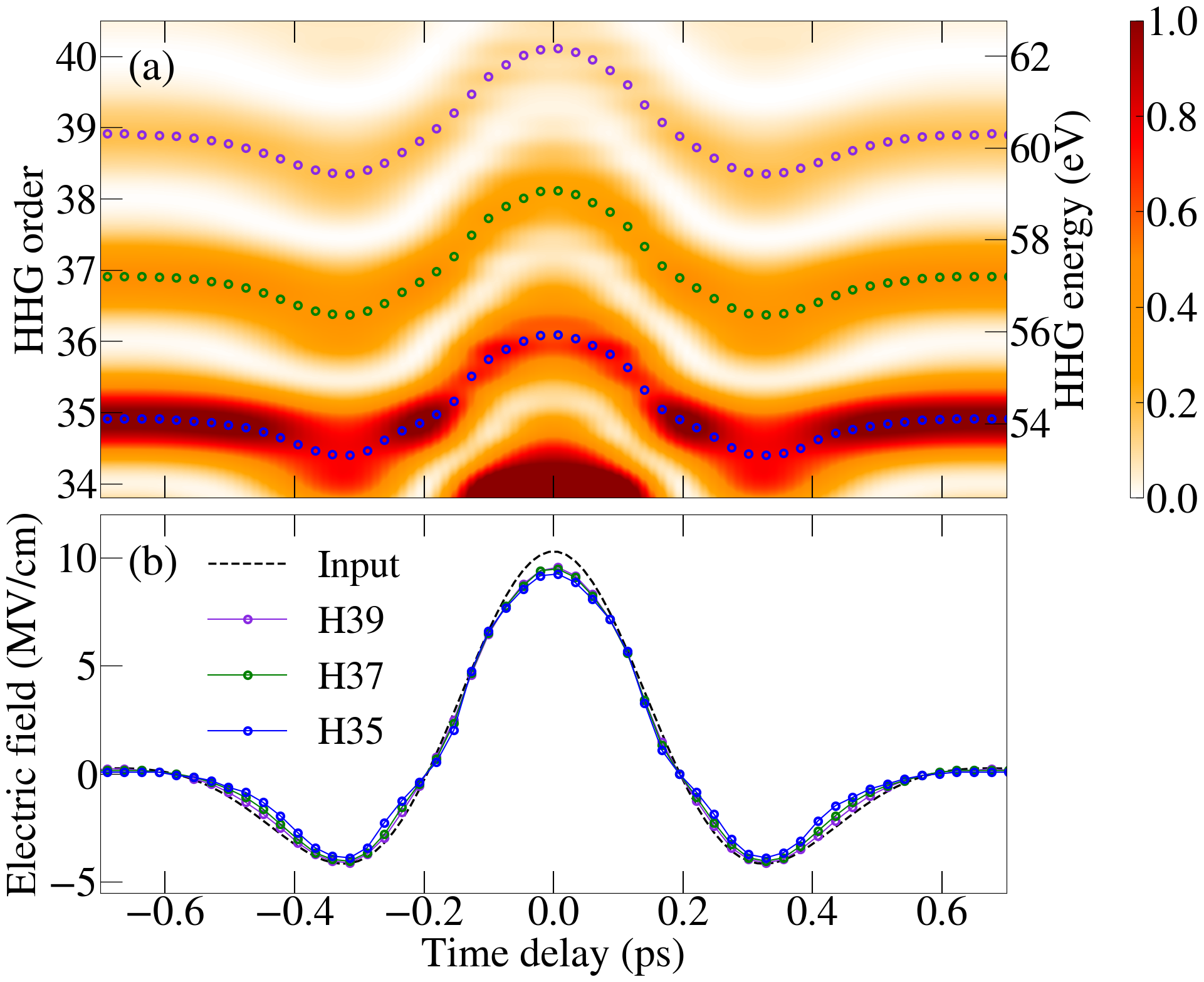}
    \caption{\textit{In situ} sampling THz waveforms~(b) by the frequency shifts of harmonics emitted with time delays between the pump (THz) and probe (few-cycle driving laser) pulses~(a). Figure~(a) displays TDSE-simulated harmonics near the cutoff with time delays, whose peak centroids are depicted by open circles. Figure~(b) shows the good consistency of the ``input'' THz pulse and those constructed from the shifts of harmonic order 35, 37, and 39 by applying analytical formula~(\ref{eq:ana}).  The used ``input'' THz pulse is $E_T(t) = E_{0T}\exp[-(\omega_T t/3\pi)^2]\cos(\omega_T t)$, where $\omega_T = 1.3$ THz, and $E_{0T}= 10.28$ MV/cm. 
    The probe pulse has the same parameters as used in Fig~\ref{fig:numerical}.
    }
    \label{fig:extract}
\end{figure}

\noindent{\textit{(2) Terahertz metrology:} } Detecting THz-waveform is an essential problem that has been intensively investigated in recent decades~\cite{Zhang:Fron21,Zhang:NatPho17}. Yet, the current array of available techniques necessitates careful consideration when selecting the appropriate active matter, underscoring the importance of developing a matter-independent route for THz detection. In Ref.~\cite{Trieu:PRA23}, we proposed a universal and target-free method for sampling THz pulses based on the intensity ratio of adjacent even and odd harmonics. However, while effective in capturing the time-dependent magnitude of THz pulses, this method falls short in suffering a $\pi$-flip uncertainty for pulse's CEP. Moreover, long-wavelength lasers are needed, so the generated harmonics near cutoff are extremely high-energy; thus, high-resolution spectrometers are required to distinguish odd and even harmonics. 
 
Using analytical relation~(\ref{eq:ana}), we propose a more direct target-free route based on a pump-probe scheme that can measure both the time-dependent magnitude and sign of a THz pulse with regular spectrometers. A THz pulse is used as a pump pulse in the pump-probe scheme. Then, delayed few-cycle laser pulses are used as probe pulses irradiating to atomic or symmetric molecular gas to generate HHG. Scanning harmonic traces gives harmonic frequency shifts as a function of time delays, as exemplified in Fig.~\ref{fig:extract}(a). Applying analytical relation~(\ref{eq:ana}), we can effectively extract both the magnitude and sign of the instantaneous THz field, see Fig.~\ref{fig:extract}(b). 

\textit{Conclusion --- } We have successfully derived a simple analytical formula for directly controlling sub-cycle electron phases -- the core of strong-field physics. The formula connects the frequency shift of harmonics near the cutoff and the low-frequency electric field when irradiated with a few-cycle infrared laser pulse into atomic or molecular gas. The formula is universal and independent of target information as long as it is centrosymmetric. The reliability of the analytical relation has been validated numerically by solving the time-dependent Schr{\"o}dinger equation.

Controlling electron phases within a sub-cycle time is essential for various \textit{in situ} applications. We have highlighted its benefits in continuously and precisely tuning the frequency of XUV waves, which is crucial in pump-probe experiments. Also, this analytical relation enables direct sampling THz pulses using regular spectrometers, irrespective of the chosen active matter.

\begin{acknowledgments}
\textit{Acknowledgments --- }
This work was funded by Vingroup and supported by Vingroup Innovation Foundation (VINIF) under project code VINIF.2021.DA00031. The calculations were executed by the high-performance computing cluster at Ho Chi Minh City University of Education, Vietnam.

\end{acknowledgments}

\bibliographystyle{apsrev4-1}
\bibliography{refs.bib}

\end{document}